\documentclass{article}

\date{\today}

\newcommand{\insertplot}[5]{\begin{figure}
 \hfill\hbox to 0.05in{\vbox to #5in{\vfill
 \inputplot{#1}{#4}{#5}}\hfill}
 \hfill\vspace{-.1in}
 \caption{#2}\label{#3}
 \end{figure}}
 \newcommand{\inputplot}[3]{
 \special{ps: plotfile #1}
}
\newcounter{fig}   

\usepackage{epsfig}
\usepackage{amsmath}
\usepackage{amsfonts}
\usepackage{graphicx}
\usepackage[german, english]{babel}
\usepackage{a4wide}
\usepackage{amsmath}
\usepackage{amssymb}
\usepackage{ifthen}
\usepackage{epsfig}

\begin{document}
\newcommand{\BA}{\ensuremath{\mathbf{A}}}
\newcommand{\Beta}{\ensuremath{\mathbf{\eta}}}
\newcommand{\BV}{\ensuremath{\mathbf{V}}}
\newcommand{\Bn}{\ensuremath{\mathbf{n}}}
\newcommand{\BW}{\ensuremath{\mathbf{W}}}
\newcommand{\BD}{\ensuremath{\mathbf{D}}}
\newcommand{\BP}{\ensuremath{\mathbf{P}}}
\newcommand{\BU}{\ensuremath{\mathbf{U}}}
\newcommand{\BR}{\ensuremath{\mathbf{R}}}
\newcommand{\BJ}{\ensuremath{\mathbf{J}}}
\newcommand{\Balpha}{\ensuremath{\boldsymbol{\alpha}}}
\newcommand{\Bbeta}{\ensuremath{\boldsymbol{\beta}}}
\newcommand{\Bphi}{\ensuremath{\boldsymbol{\phi}}}
\newcommand{\Btau}{\ensuremath{\boldsymbol{\tau}}}
\newcommand{\Bpi}{\ensuremath{\boldsymbol{\pi}}}
\newcommand{\Bxi}{\ensuremath{\boldsymbol{\xi}}}
\newcommand{\Br}{\ensuremath{\mathbf{r}}}
\newcommand{\Bx}{\ensuremath{\mathbf{x}}}
\newcommand{\Bq}{\ensuremath{\mathbf{q}}}
\newcommand{\Bp}{\ensuremath{\mathbf{p}}}
\newcommand{\Bv}{\ensuremath{\mathbf{v}}}
\newcommand{\Bh}{\ensuremath{\mathbf{h}}}
\newcommand{\BH}{\ensuremath{\mathbf{H}}}
\newcommand{\Ba}{\ensuremath{\boldsymbol{a}}}
\newcommand{\BPhi}{\ensuremath{\boldsymbol{\Phi}}}
\newcommand{\Bell}{\ensuremath{\boldsymbol{\ell}}}
\newcommand{\Bsigma}{\ensuremath{\boldsymbol{\sigma}}}

\newcommand{\Z}{\ensuremath{\mathds{Z}}}
\newcommand{\NN}{\ensuremath{\mathds{N}}}
\newcommand{\R}{\ensuremath{\mathds{R}}}
\newcommand{\SU}{SU(2)}
\newcommand{\su}{{su(2)}}
\newcommand{\e}{\mathrm{e}}
\newcommand{\rt}{\tilde{r}}
\newcommand{\di}{\mathrm{d}}
\newcommand{\dd}{\mathrm{d}}
\newcommand{\ii}{\mathrm{i}}
\newcommand{\CH}{\ensuremath{\mathcal{H}}}
\newcommand{\CL}{\ensuremath{\mathcal{L}}}
\newcommand{\CB}{\ensuremath{\mathcal{B}}}
\newcommand{\CM}{\ensuremath{\mathcal{M}}}
\newcommand{\CG}{\ensuremath{\mathcal{G}}}
\newcommand{\CE}{\ensuremath{\mathcal{E}}}
\newcommand{\CA}{\ensuremath{\mathcal{A}}}
\newcommand{\CF}{\ensuremath{\mathcal{F}}}
\newcommand{\be}{\begin{equation}}
\newcommand{\bea}{\begin{eqnarray}}
\newcommand{\tr}{\mbox{tr}}
\newcommand{\la}{\lambda}
\newcommand{\ta}{\theta}
\newcommand{\f}{\phi}
\newcommand{\vf}{\varphi}
\newcommand{\ka}{\kappa}
\newcommand{\al}{\alpha}
\newcommand{\ga}{\gamma}
\newcommand{\de}{\delta}
\newcommand{\si}{\sigma}
\newcommand{\bomega}{\mbox{\boldmath $\omega$}}
\newcommand{\bsi}{\mbox{\boldmath $\sigma$}}
\newcommand{\bchi}{\mbox{\boldmath $\chi$}}
\newcommand{\bal}{\mbox{\boldmath $\alpha$}}
\newcommand{\bpsi}{\mbox{\boldmath $\psi$}}
\newcommand{\brho}{\mbox{\boldmath $\varrho$}}
\newcommand{\beps}{\mbox{\boldmath $\varepsilon$}}
\newcommand{\bxi}{\mbox{\boldmath $\xi$}}
\newcommand{\bbeta}{\mbox{\boldmath $\beta$}}
\newcommand{\ee}{\end{equation}}
\newcommand{\pa}{\partial}
\newcommand{\Om}{\Omega}
\newcommand{\vep}{\varepsilon}
\newcommand{\bfph}{{\bf \phi}}
\newcommand{\eea}{\end{eqnarray}}

\title{Baby Skyrmions stabilized by canonical quantization}

\author{
{\large A.Acus}$^{\dagger}$, {\large E. Norvai\v{s}as}$^{\dagger}$
and {\large Ya. Shnir}$^{\ddagger \star}$ \\ \\
\\ $^{\dagger}${\small Vilnius University Institute of Theoretical Physics and Astronomy,}
\\ {\small Go\v{s}tauto 12, Vilnius 01108, Lithuania}
\\ $^{\ddagger}${\small School of Theoretical Physics -- DIAS, 10
Burlington Road, Dublin 4, Ireland  }
\\$^{\star}${\small Department of Computer Science, National
University of Ireland Maynooth}}

\maketitle

\begin{abstract}
We analyse the effect of the canonical quantization of the rotational mode of the $O(3)$ $\sigma$-model
which includes the Skyrme term.  Numerical evidence is presented that
the quantum correction to the mass of the rotationally-invariant charge $n=1,2$ configurations may
stabilize the solution even in the limit of vanishing potential. The corresponding range of values of the
parameters is discussed.
\end{abstract}


\section{Introduction}

In this paper we are concerned with canonical quantization of the
soliton solutions of the modified version of the $d=2+1$
$O(3)$ $\sigma$-model which includes the Skyrme term \cite{Piette:1994jt,Piette:1994ug}
(so-called baby Skyrme model). The model can be viewed as a lower
dimensional simplified analogue of the usual Skyrme model \cite{Skyrme:1961vq}, however it
has its own physical application in condensed matter physics where the soliton solutions
of the model describe spin textures in the ferromagnetic quantum Hall system  (see, e.g.,
\cite{Sondhi:1993zz,Walet:2001zz,Ezawa:2000ae}).

Unlike the usual Skyrme model in $d=3+1$, a potential term is
usually added to the classical baby Skyrme model to ensure stability of the skyrmions
in two-dimensional space by the Derrick scaling scaling requirements. 
On the other hand, in the two-dimensional ferromagnetic system such
a term describes the coupling with an external magnetic field.

In Skyrme's original model which is considered as
an effective theory of low-energy QCD in the limit of large number of quark colours
\cite{Witten:1983tw}, this potential term is optional, it corresponds to the no-zero pion mass although its
appearance might dramatically affect the structure of the solutions \cite{Battye:2004rw}.
The form of the potential term in the baby Skyrme model is largely arbitrary, there are
different families of possible models, e.g., holomorphic model which yields a simple
analytic solution written in terms of the holomorphic functions \cite{Leese:1989gi}, or
double vacuum model which gives rise to the circular multiskyrmions \cite{Weidig:1998ii}.
Actually the choice of potential term of the baby Skyrme model
dictates the structure of the multi-skyrmions
\cite{Hen:2007in,Karliner:2009at}. Note that apart inclusion of the potential term,
there are other possibilities to ensure stability of the soliton solutions,
for example by coupling of the $O(3)$ sigma-model to a massive vector meson field \cite{Foster:2009rw}.

Standard approach to the quantization of the Skyrmions involves the
zero-mode quantization of the configuration as a rigid body \cite{Adkins:1983hy,Adkins:1983ya}.
Furthermore, the classical Skyrmion is quantized within the Bohr-Sommerfeld framework by requiring the angular
momentum to be quantized, i.e., the quantum excitations correspond to a spinning Skyrmion
with a particular rotation frequency. In the recent paper \cite{Battye:2005nx}
an axially symmetric ansatz was used to allow the spinning
Skyrmion to deform, it was shown that the standard Skyrme parameters are simply
an artifact of the rigid body approximation.
The rotating solutions of the baby Skyrme model are also known \cite{Piette:1994mh,Betz:1996rh},
in this case the rotationally symmetric configuration is stabilized due to radiative effects.

Furthermore, it was suggested to treat the Skyrme model
quantum mechanically, i.e., apply  canonical quantization of the collective coordinates of
the soliton solution to take into account quantum mass corrections
\cite{Fujii:1986wt,Fujii:1986fu,Acus:1996um,Jurciukonis:2005em}. It turns out the correction
decreases the mass of the configuration, so one can expect similar effect in the two-dimensional
baby Skyrme model.

In this paper we observe numerically that the canonical quantization of the rotational degrees of freedom of the
baby Skyrme model produces a stable soliton solution even in the limit of vanishing potential term.
The stabilization is achieved by including of the
corresponding quantum corrections which appear when the canonical commutation relations are taken into account
and decrease the mass of the soliton.

\section{The baby Skyrme model}

Let us begin with a brief review of the Skyrme model in 2+1 dimensions.
In terms of the order parameter triplet $\Bphi=(\phi_{1},\phi_2,\phi_3)$, of the
nonlinear $O(3)$ sigma model field subject to $|\phi_a|^2=1$, the Lagrangian density of the
baby Skyrme model has the form
\begin{eqnarray}
{\cal L} = \frac{\alpha^2}{2} \partial_\mu \Bphi \cdot \partial^\mu \Bphi - \frac{\kappa^2}{4}
(\partial_\mu \Bphi \times \partial_\nu \Bphi)^2 - \mu^2(1- \phi_3).
\label{LagrangianDensity1}
\end{eqnarray}

The condition of
finiteness of the energy requires the potential term to vanish at infinity. This implies that
on the spacial boundary $\Bphi \to (0,0,1)$. Therefore the physical space $\mathbb{R}^2$ is compactified to
$S^2$ and such  a one-point compactification ensures the topologically non-trivial mapping
$\Bphi : S^2 \to S^2$. The corresponding topological charge is the homotopy invariant
\be \label{charge}
n =  \varepsilon_{abc}\frac{1}{4\pi}\int\limits_{\mathbb{R}^2}\mathrm{d}^2x \phi_a\partial_1 \phi_b\partial_2 \phi_c
\ee

There are three free parameters in the model.
The constants $\alpha^2$, $1/\kappa^2$, $\kappa^2 \times [length]^{-2}$ and $\mu^2 \times  [length]^2$
have the dimension of energy. On the classical level the energy scale provided by the parameter
$\alpha^2$ can be fixed in geometric units by setting $\alpha^2 = 1$, furthermore, in that case
the length scale also can be fixed by setting $\kappa^2$ to 1 but the rescaled classical parameter $\mu$
remains free \cite{Piette:1994mh}. In natural units where the Planck constant is normalized
to unity  we can introduce dimensionless radial coordinate
$\rho=\frac{\alpha}{\kappa}r$. Then the energy functional of the static field configurations
can be conveniently written in polar coordinates
$r, \theta$ or in dimensionless coordinates $\rho, \theta$ as
\begin{eqnarray}
E = \int r \mathrm{d}r \mathrm{d}\theta \left(\frac{\alpha^2}{2} [\partial_r \Bphi \cdot  \partial_r \Bphi + \frac{1}{r^2}
 \partial_\theta \Bphi \cdot  \partial_\theta \Bphi ] + \frac{\kappa^2}{2 r^2} (\partial_r \Bphi \times
\partial_\theta \Bphi)^2 + \mu^2 (1-\phi_3)
 \right)\\
 = \frac{\alpha^2}{2}\int \rho \mathrm{d}\rho \mathrm{d}\theta \left( \partial_\rho \Bphi \cdot  \partial_\rho \Bphi + \frac{1}{\rho^2}
 \partial_\theta \Bphi \cdot  \partial_\theta \Bphi  + \frac{1}{\rho^2} (\partial_\rho \Bphi \times
\partial_\theta \Bphi)^2 + \frac{2\mu^2}{\alpha^2} (1-\phi_3)
 \right)\, ,
\end{eqnarray}
respectively.
However, when the quantum corrections to the soliton mass
are taken into account, the energy scale is changing. Thus we cannot set $\alpha^2 = \kappa^2 = 1$ anymore and,
as we will see both parameters are significant in that case.

In this paper we only concerned with rotationally-invariant fields, so
the corresponding $O(2)$ symmetry of the system  is commonly used to re-express the field  $\Bphi$
in terms of a profile function $f(r)$ or $f(\rho)$, and a polar angle $\theta$ as
\begin{equation}
\Bphi=
\Bigl(
 \sin f(r)\cos n \theta,\quad
 \sin f(r)\sin n \theta,\quad
 \cos f(r)
\Bigr),
\label{vectorAnsatz}
\end{equation}
where we set the global phase to be zero. $f(r)$ is the real profile function
which satisfies certain boundary conditions. Here we take $f(0) = \pi$ and $f(\infty) = 0$.
The integer $n$ is actually the topological charge
of the configuration, as one can see substituting the ansatz \eqref{vectorAnsatz} into the definition
\eqref{charge}. For the topological sectors with $n=1$ and $n=2$ this parametrization
provides remarkably accurate approximation to the exact numerical solution of the
model~\eqref{LagrangianDensity1} \cite{Piette:1994ug}. For higher values of the topological charges $n > 3$
this approach yields unstable circular multisoliton configurations, so the ground state solutions are not
rotationally-symmetric although the structure of the solution depends on the explicit form of the potential
of the model \cite{Hen:2007in,Karliner:2009at,Foster:2009vk}

In order to apply the standard  canonical quantization procedure it is convenient to re-express the Lagrangian
\eqref{LagrangianDensity1} in terms of the $SU(2)$-valued hermitian matrix fields $U = \Bphi \cdot \tau$,  where
$\tau = (\tau_1,\tau_2, \tau_3)$ is the triplet of usual Pauli matrices
\begin{equation}
\label{matrixAnsatz}
U(f(r),\theta)=
\left(
\begin{array}{ll}
 \cos f(r) &\sin f(r) \mathrm{e}^{-\mathrm{i} n \theta }  \\
 \sin f(r) \mathrm{e}^{\mathrm{i} n \theta } & -\cos f(r)
\end{array}
\right).
\end{equation}
The asymptotic value of the field $U(f(r),\theta)$ has to tend to the matrix $\tau_3$.

Then the Lagrangian~\eqref{LagrangianDensity1} can be represented in the form similar to the usual
structure of the Skyrme model in $d=3+1$
\begin{equation}\label{LagrangianDensity2}
\begin{split}
{\cal L} &=
 \frac{\alpha^2}{4} \mathop{\mathrm{Tr}} \partial_k U \partial^k U + \frac{\kappa^2}{32}
\mathop{\mathrm{Tr}}\bigl[ \partial_k U, \partial_l U \bigr]\bigl[ \partial^k U, \partial^l U \bigr]
-\frac{\mu^2}{2} \mathop{\mathrm{Tr}}\bigl(\mathbf{1}-\tau_3 U\bigr),
\end{split}
\end{equation}

Substituting the ansatz~\eqref{matrixAnsatz} into the classical Lagrangian density
\eqref{LagrangianDensity2} gives
\begin{equation}
\label{ClassicalBabyLag}
\begin{split}
-\CL_{cl}=&\frac{\alpha^2}{2}f^{\prime2} + \frac{n^2\sin^2 f}{2r^2}(\alpha^2+\kappa^2 f^{\prime2})
+\mu^2\bigl(1-\cos f\bigr).
\end{split}
\end{equation}
The Lagrange density \eqref{ClassicalBabyLag} leads to the
classical Euler-Lagrange equation
\begin{equation}
\begin{split}
&f^{\prime\prime}\Bigl(\alpha^2r + \frac{\kappa^2 n^2 \sin ^2 f}{r}\Bigr)+ f^{\prime}\Bigr(\alpha^2 +
f^\prime \frac{\kappa^2 n^2\sin f \cos f
 }{r} - \frac{\kappa^2 n^2 \sin ^2 f}{r^2}\Bigr)\label{ClassicalBabyEq}\\
&~~~~~~~~~~~~~~~~~~~~~~~~~~~
-\frac{\alpha^2n^2 \sin f \cos f }{r}-\mu^2r \sin f=0,
\end{split}
\end{equation}
which can be integrated numerically subject of the boundary conditions imposed \cite{Piette:1994ug}.
The equation \eqref{ClassicalBabyEq} in the dimensionless coordinates takes the form
\begin{equation}
\begin{split}
&f^{\prime\prime}\Bigl(\rho + \frac{n^2 \sin ^2 f}{\rho}\Bigr)+ f^{\prime}\Bigr(1 +
f^\prime \frac{n^2\sin f \cos f}{\rho}
- \frac{n^2 \sin ^2 f}{\rho^2}\Bigr)\label{ClassicalBabyEqDimles}\\
&~~~~~~~~~~~~~~~~~~~~~~~~~~~
-\frac{n^2 \sin f \cos f }{\rho}-\frac{\kappa^2\mu^2}{\alpha^4}\rho \sin f=0.
\end{split}
\end{equation}
Thus the stability of static soliton solutions depends on a single free dimensionless
parameter $\frac{\kappa^2\mu^2}{\alpha^4}$.

\section{Quantization: Momenta of inertia}

We wish to quantize the rotational degrees of freedom of baby Skyrmion
by wrapping classical baby
Skyrmion ansatz $U \bigl(f(r),\theta \bigr)$ with unitary matrices $\BA\bigl(\Bq(t)\bigr)$
depending only on time $t$ \cite{Adkins:1983ya}
\begin{equation}
\BU(\Bq,f,\theta)=\BA\bigl(\Bq(t)\bigr) U \bigl(f(r),\theta\bigr) \BA^\dagger\bigl(\Bq(t)\bigr).
\end{equation}
Here, for the sake of generality, we suppose the field~\eqref{vectorAnsatz} of the model is embedded into the 3-dimensional isospace.
Then the three Euler angles are associated with collective rotational degrees of freedom $\Bq(t)$
will eventually be treated as quantum-mechanical variables.
The generalized coordinates $\Bq(t)$ and velocities $\dot \Bq(t)$ then satisfy the commutation
relations \cite{Fujii:1986fu}
\begin{equation}
\lbrack \dot q^a,\,q^b]=-\mathrm{i}f^{ab}(\Bq).
\label{CommutationRelation1}
\end{equation}
The explicit form of the function $f^{ab}(\Bq)$ will be completely determined by canonical
commutation relations between quantum coordinates and momenta.

As usual, to calculate the effective Lagrangian of the rotational zero mode we have to
evaluate the time derivative of the matrix
\begin{equation}
\dot \BU=\dot \BA U \BA^\dagger - \BA U \BA^\dagger\dot \BA \BA^\dagger,
\end{equation}
taking into account the commutation relations~\eqref{CommutationRelation1}. Explicitly, we have
\begin{equation}
\dot \BA\bigl(\Bq(t)\bigr)=\frac12 \bigl\{\dot q^k,\frac{\partial}{\partial q^k} \BA\bigr\},
\label{Anticommutator}
\end{equation}
where
\begin{equation}
\frac{\partial}{\partial q^k} \BA(\Bq)= C^{(a)}_k(q)\tau_a \BA(\Bq)= C^{\prime (a)}_k(\Bq)\BA(\Bq) \tau_a,
\end{equation}
and the curly brackets in~\eqref{Anticommutator} correspond to the anticommutator.
The coefficients $C^{(a)}_k, C^{\prime (a)}_k$ are some functions of the group parameters whose explicit form
is not relevant here.

Then, keeping only terms proportional to the square of the angular velocity
in the effective kinetic Lagrangian density, we get
\begin{equation}
\begin{split}
\mathcal{L}_q(f(r),\theta)&=\frac12 \dot\Bq ^k g_{k,k^\prime} \dot\Bq ^{k^\prime} +\cdots =\frac12 \dot\Bq ^k C^{\prime (a)}_k (\Bq ) \mathcal{E}_{a,b}(f(r),\theta) C^{\prime (b)}_{k^\prime} (\Bq ) \dot\Bq ^{k^\prime}
 +\cdots
\end{split}
\end{equation}
where the metric of the restricted configuration space of the rotational zero modes is
\begin{equation}
g_{k,k^\prime}(q,f(r),\theta)=C^{\prime (b)}_k (q) \mathcal{E}_{b b^\prime}(f(r),\theta) C^{\prime (b^\prime)}_{k^\prime} (q).
\end{equation}
Performing explicit summation in the circular basis $\tau_{+}=-\frac{1}{\sqrt{2}}(\tau_1 + \mathrm{i}\tau_2)$, $\tau_{0}=\frac{1}{2}\tau_3$, $\tau_{-}=\frac{1}{\sqrt{2}}(\tau_1 - \mathrm{i}\tau_2)$ we find
\begin{align*}
&&\begin{split}
&\mathcal{E}_{b b^\prime}(f(r),\theta)= \frac{\alpha^2}{4}
\left(
\begin{array}{lll}
\sin ^2f \mathrm{e}^{2 \mathrm{i} n\theta }  & -\sin 2f \frac{\mathrm{e}^{\mathrm{i} n\theta } }{\sqrt{2}} & 2-\sin ^2f \\
 - \sin2f \frac{\mathrm{e}^{\mathrm{i} n\theta }}{\sqrt{2}} & -2 \sin ^2f &  \sin2f\frac{\mathrm{e}^{-\mathrm{i} n\theta }}{\sqrt{2}} \\
 2-\sin ^2f & \sin2f \frac{\mathrm{e}^{-\mathrm{i} n\theta }}{\sqrt{2}} &  \sin ^2f \mathrm{e}^{-2 \mathrm{i} n\theta }
\end{array}
\right)
\\
&
+\frac{\kappa^2}{4}
 \left(
\begin{array}{lll}
 -\mathrm{e}^{2 \mathrm{i} n\theta } \left(\cos ^2f f^{\prime2}-\frac{n^2\sin ^2f}{r ^2}\right)
& -\frac{\mathrm{e}^{\mathrm{i} n\theta } f^{\prime2} \sin 2 f }{\sqrt{2}}
& \cos ^2f f^{\prime2}+\frac{n^2\sin ^2f}{r^2} \\
 -\frac{\mathrm{e}^{\mathrm{i} n\theta } f^{\prime2} \sin 2 f}{\sqrt{2}} & -2 f^{\prime2} \sin ^2f &
 \frac{\mathrm{e}^{-\mathrm{i} n\theta } f^{\prime2} \sin 2 f}{\sqrt{2}}\\
 \cos ^2 f f^{\prime2}+\frac{n^2\sin ^2 f}{\rho ^2}
& \frac{\mathrm{e}^{-\mathrm{i} n\theta } f^{\prime2} \sin 2 f}{\sqrt{2}}
&  -\mathrm{e}^{-2 \mathrm{i} n\theta } \left(\cos ^2 f f^{\prime2}-\frac{n^2\sin ^2 f}{r^2}\right)
\end{array}
\right).
\end{split}
\end{align*}
Integration of the corresponding matrix
\begin{equation}
\label{Ematrix}E_{b b^\prime}=\int r \dd r \dd \theta \mathcal{E}_{b b^\prime}(f(r),\theta) =
\begin{pmatrix}
0 & 0 &a_1\\
0 &  -a_0 & 0\\
a_1 & 0 &0
\end{pmatrix}
\end{equation}
gives the explicit expressions for the baby Skyrmion's momenta of ``inertia''
\begin{equation}
\label{momentaOfInertia}
a_0=\pi \int_0^\infty r \sin^2 f \Bigl(\alpha^2 + \kappa^2 f^{\prime 2} \Bigr)\,\mathrm{d} r
=\kappa^2\pi \int_0^\infty \rho \sin^2 f \Bigl(1 + f^{\prime 2} \Bigr)\,\mathrm{d} \rho
\equiv \kappa^2 \tilde{a}_0,
\end{equation}
\begin{equation}
a_1=\frac{\pi}{2} \int_0^\infty r \Bigl( \alpha^2\bigl(\sin^2 f -2\bigr) +\kappa^2\bigl(f^{\prime 2} \cos^2 f
+ \frac{n^2\sin^2 f}{r^2}\bigr)\Bigr)\,\mathrm{d} r .
\label{momentaOfInertia1}
\end{equation}

Evidently, the integral $a_1$ in~\eqref{momentaOfInertia1} contains a divergent term.
Physically, it means the rotations around corresponding axes are forbidden as expected.
The moment of inertia $a_0$ of a spinning baby Skyrmion, however,
diverges only in the limit when angular velocity of the rotation $q$ approaches the
value of $\mu$ \cite{Piette:1994mh} and generally
the rotations around the 3rd axis are allowed. Restricting ourselves to the rotations about
this axis, we fix the U(1) subgroup $\BA(\Bq)=\exp(\mathrm{i}q\tau_3/2)=
\cos \frac{q}{2} \mathbf{1}+\mathrm{i}\sin \frac{q}{2}\tau_3$
of the complete isospin rotation group.

\begin{figure}
\begin{center}
\includegraphics*[scale=0.45,keepaspectratio]{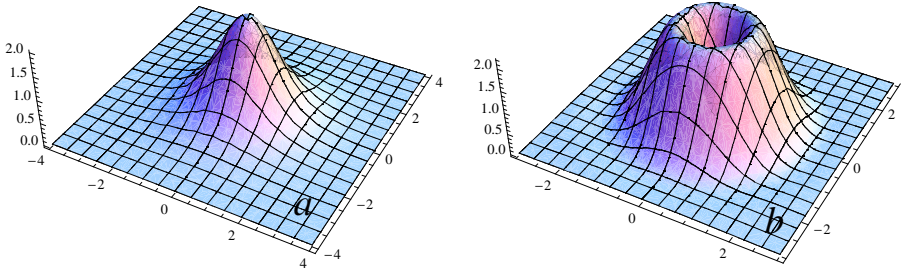}
\caption{A typical energy density distribution for winding number $n=1$~($a$) and $n=2$~($b$) sectors.
Parameters of the plotted $n=1$ solution are $\varepsilon^2=28$, $\mu_s=0$, $\omega^2=0$ and correspondingly $\varepsilon^2=57$, $\mu_s=0$, $\omega^2=0$ for the $n=2$ sector.}
\label{constantHeSurface}
\end{center}
\end{figure}

Explicit calculation then gives $\BA^\dagger (q)\bigl(\frac{\partial}{\partial q}
\BA(q)\bigr)=\frac{\mathrm{i}}{2}\tau_3$, $C(q)=C^{\prime}(q)=\mathrm{i}$, and only the
middle entry of the  matrix  \eqref{Ematrix} remains.

We are interested in a quantum-mechanical model where the corresponding quantum momentum
is  conjugated to the rotational collective coordinate $q$ and it is defined as
\begin{eqnarray}
\hat{p} &=&\frac{\partial L_q}{\partial \dot{q}}=\dot q a_0 .
\label{momentum}
\end{eqnarray}
The canonical commutation relation $\lbrack \hat{p},\,q^b]=-\mathrm{i}$
the allows us to define the explicit form of the algebra ~\eqref{CommutationRelation1}
\begin{eqnarray}
f^{00}(q) &=&\frac{1}{a_0}.
\label{ffunction}
\end{eqnarray}

\section{Quantum Lagrangian and equation of motion}

We are now in position to evaluate the explicit form of the quantum-mechanical
Lagrangian of the baby Skyrme model. Simple calculation yields
$\BA^\dag \dot \BA =\frac{\mathrm{i}}{2} \tau_3 \dot q +\frac{\mathrm{i}}{8 a_0}\mathbf{1}$
and the quantum Lagrangian is given by
\begin{equation}
L_q=\int \mathcal{L}_q[f (r)]=\frac{\hat p^2}{2 a_0}-\Delta M
\end{equation}
where
\begin{eqnarray}
\Delta M=
-\frac{\pi}{8a^2_0}\int r \mathrm{d} r \biggl( \alpha^2\sin^2 f
+\frac{\kappa^2}{32} \Bigl(32 f^{\prime 2} \sin^2 f+f^{\prime 2}
+\frac{n^2\sin^2 f}{r^2} - \frac{2 n^2\sin^4 f}{r^2}\Bigr)\biggr)\\
=
-\frac{\pi}{8 \kappa^2 \tilde{a}^2_0}\int \rho \mathrm{d} \rho \biggl( \sin^2 f
+\frac{1}{32} \Bigl(32 f^{\prime 2} \sin^2 f+f^{\prime 2}
+\frac{n^2\sin^2 f}{\rho^2} - \frac{2 n^2\sin^4 f}{\rho^2}\Bigr)\biggr)=
\frac{\Delta \tilde {M}}{\kappa^2}
\end{eqnarray}
is the quantum mass correction which appear when the commutation relations~\eqref{CommutationRelation1}
are taken into account.

We define the angular momentum operator as
\begin{eqnarray*}
\hat J &=&-\frac{\mathrm{i}}{2}\bigl\{\hat p,C^{-1}(q)\bigr\}=- a_0 \dot q.
\end{eqnarray*}
Its eigenstates are the vectors
\begin{eqnarray*}
|\omega \rangle=\mathop{\mathrm{exp}}(-\mathrm{i} \omega q)|0\rangle,
\end{eqnarray*}
where $\omega$ is an integer which enumerates the irreducible representations of the U(1) group.

The total effective Hamiltonian corresponds to the complete Lagrangian $L=L_{cl}+L_q$ which includes
both classical and quantum mechanical parts:
\begin{equation}
H=\frac{1}{2}\{\hat p,\dot q\}-L=\frac{{\hat J}^2}{2 a_0}-L_{cl}+\Delta M,
\label{hamiltonian}
\end{equation}

In the framework of the Bohr-Sommerfeld quantization of the rotational zero mode of the
baby Skyrmion, the quantum mass term is absent \cite{Piette:1994mh}. This picture corresponds to
the rigid rotation of the Skyrmion with fixed profile of the function $f(r)$.
However, since the quantum mass correction $\Delta M$ turns out to be negative it might
stabilize the baby Skyrmion solutions as it happens
in the three-dimensional Skyrme model \cite{Acus:1997za}. This is the most interesting problem
to study, so henceforth we shall mainly consider the limiting case of vanishing potential term $\mu \to 0$.

Indeed, let us consider the equations which corresponds to the
minimization of the total energy functional. Varying it we obtain rather cumbersome
integro-differential equation in dimensionless coordinates on the profile function $f(\rho)$
replacing its classical counterpart
\eqref{ClassicalBabyEq}:
\begin{equation}
\label{QuantumEquation}
\begin{split}
&f^{\prime\prime}(\rho)
\left(\frac{2 n^2\sin ^2f(\rho)}{\rho} + 2\rho \bigl(1+  {\mathbb{Z}} \mu_d^2 \sin^2 f(\rho) \bigr)
\right)
\\
+&
f^{\prime 2}(\rho)\left(\frac{n^2\sin 2f(\rho)}{\rho} + \rho  {\mathbb{Z}} \mu_d^2 \sin 2 f(\rho)
\right)\\
+&
f^{\prime}(\rho)\left(-\frac{2 n^2\sin ^2f(\rho)}{\rho^2} + 2  {\mathbb{Z}}
\bigl(1+ \mu_d^2\sin^2 f(\rho) \bigr)
\right)
\\
-&\sin 2 f(\rho) \left(\rho  {\mathbb{Z}} \mu_d^2
+
\frac{n^2}{\rho}
( {\mathbb{Z}} + 4 \sin^2 f(\rho)(1- {\mathbb{Z}}) )
\right)
\\
-&2  {\mathbb{Z}} \mu_s^2 \rho \sin f(\rho)
= 0
\end{split}
\end{equation}

Here the usual boundary conditions on the function $f(\rho)$ are imposed and
we introduce the shorthand notations for the dimensionless quantities
\begin{equation}
\begin{split}
& {\mathbb{Z}} = 1- \frac{1}{2^8 \alpha^2 \kappa^2 \tilde{a}_0^2};\qquad
\mu_s^2= \frac{\kappa^2 \mu^2}{\alpha^4 {\mathbb{Z}}}
\end{split}
\end{equation}
and
\begin{equation}
\begin{split}
&\mu_d^2= \frac{1}{2 {\mathbb{Z}} \alpha^2\kappa^2 \tilde{a}_0^2}\Bigl(\frac14 - \omega^2 +
\frac{I_B}{2^6 \tilde{a}_0}\Bigr)
\end{split}
\end{equation}

where
\begin{equation}
I_B=\pi \int_0^\infty \Bigl( \rho f^{\prime 2}(\rho) +\frac{n^2\sin^2f(\rho)}{\rho} -
\frac{2n^2\sin ^4f(\rho)}{\rho}\Bigr) \mathrm{d} \rho .
\end{equation}

As $\rho \rightarrow \infty$, the
equation~\eqref{QuantumEquation} reduces to the asymptotic form
\begin{equation}
\label{asympEq}
\begin{split}
&\rho f^{\prime\prime}(\rho)
+ f^{\prime}(\rho)
- m^2 \rho f(\rho)  = 0.
\end{split}
\end{equation}
where the quantity $ m^2 = \mu_s^2 + \mu_d^2$ corresponds the asymptotic mass
of the excitations. Thus the leading term in an
asymptotic expansion of the function $f(\rho)$ is given by
\begin{equation}
\label{AsymptoticSolution}
\begin{split}
&f(\rho)= C_1
\frac{e^{-m \rho}}{\sqrt{\rho}}\Bigl(1-\frac{1}{8m\rho}\Bigr).
\end{split}
\end{equation}
and the solution of the quantum-mechanical model remains exponentially
localised. The constant of integration $C_1$ which appears here will be
determined later from the results of the numerical calculations.

\section{Numerical Results}

The integro-differential equation \eqref{QuantumEquation} can be
solved numerically by shooting method.
The initialization of the algorithm requires trial values for all the integrals
${\tilde a}_0$, $\mu_d^2$ and $I_B$ that appear in the equation to be specified.
Estimates of these can be obtained by employment of the
classical profile function $f(\rho)$ of the corresponding winding number $n$ baby Skyrmion.
If we drop out the explicit mass term $\mu_s=0$
(the most intriguing case), the stable soliton solution can be obtained only for
some range of the values of the parameter $1/\varepsilon^2=\alpha^2\kappa^2$.
This can easily be seen by noticing that in the limit $\varepsilon^2 \rightarrow 0$
the quantum equation \eqref{QuantumEquation} is formally
reduced to the classical equation~\eqref{ClassicalBabyEqDimles}.
However in this limit the classical chiral function $f(\rho)$
decays as $\sim\frac{1}{\rho}$ and, consequently the moment of inertia
given by the integral $\tilde{a}_0$ \eqref{momentaOfInertia} diverges \cite{Piette:1994mh,Betz:1996rh}.
Physically it means that as $\varepsilon^2$ decreases, the quantum
baby Skyrmion slows down and the quantum correction to the soliton solution
becomes negligible, so they cannot stabilize the configuration.

On the other side, the limit of the large values of the parameter
$\varepsilon^2$,  i.e., the case of relatively large quantum corrections,
implies small values of the quartic stabilizing term, so in this limit the
soliton solution becomes unstable. Thus, we may expect existence of a window of values of
parameter $\varepsilon^2$ for which we may get a quantum-mechanical rotated solution
with a non-vanishing potential term $\mu_s\neq 0$.

The solution of integro-differential equation then proceeds as follows.
Certainly, for some intermediate range of values of the parameter
$\varepsilon^2$, the values of the
integral  $\tilde{a}_0$ can be obtained by introducing an integration cutoff. These approximated
values then can be used
as an input for the next step of numerical iteration over the entire range of values of the
radial variable.

Shooting from the point $\rho_\textrm{max}$  (where $f(\rho)$ assumed to be of the form
\eqref{AsymptoticSolution}), to the point $\rho_\textrm{min}$
(here $f(\rho)= f(\rho_\textrm{min})- (\rho_\textrm{min}-\rho)
 f^\prime(\rho_\textrm{min})$) and varying the only unknown
constant $C_1$ in \eqref{AsymptoticSolution} yields a continuous family of solutions,
which satisfies the required topological boundary conditions $f(0)=\pi$ and $f(\infty)=0$.
Typically,  $\rho_\mathrm{max}\in [5, 12]$ and we set $\rho_\mathrm{min} \approx 10^{-3}$
to obtain solutions with topological numbers $n=1,2,3$.

For larger values of the topological charges $n\ge 4$ the  multisoliton solution
profile function $f$ slowly varies in the vicinity of the origin, so the energy density distribution
is getting more extended \footnote{Remind that the rotational symmetry of the classical baby Skyrmion solution
holds in the sectors with $n=1,2$ providing an absolute minimum of energy functional there. However
the effect of rotation may affect the structure of the discrete symmetry
solutions with higher values of the topological charge $n\ge 3$ increasing degree of symmetry up to
rotationally invariant ansatz \eqref{vectorAnsatz}. Although we do not address this issue here, we
hope to investigate this transition in our future work.}.
Technically it means the value of
$\rho_\mathrm{min}$ should be increased up to $\approx 10^{-1}$ to keep the
numeric algorithm stable.

Once the profile function is found, it can be used as an input for the next step of the iteration
procedure. Then
we recalculate all required integrals again and repeat the same procedure until all the integrals do
converge to some stable values.

The analytical and numerical calculations are performed with Mathematica~\cite{Wolfram:2007}.
Typically 20--60 shoots are enough to achieve high precision numerical solution of the
differential equation and 1000-5000 iterations is needed to ensure all the integrals are
definitely converge to some set of fixed values in the case when we set $\mu_s=0$.

\begin{figure}
\begin{center}
\includegraphics*[scale=0.45,keepaspectratio]{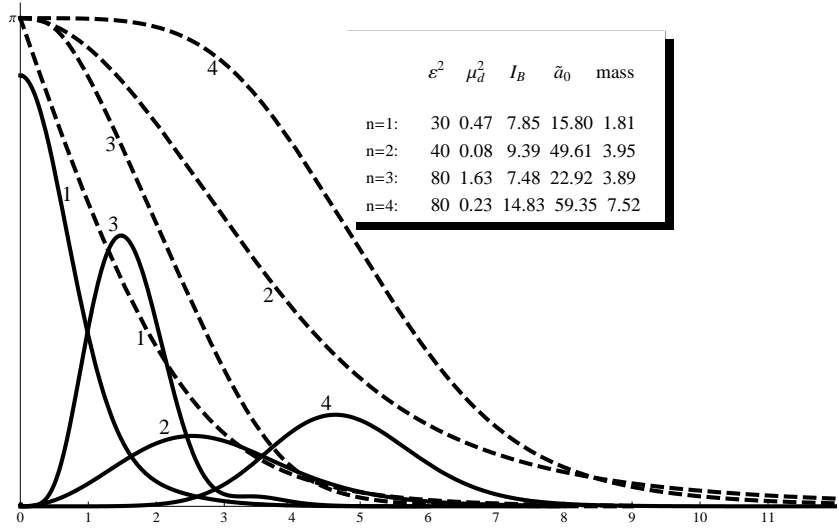}
\caption{Typical profile functions (dashed lines ) and the
distribution of the  energy density (solid lines ) of the
soliton solutions with winding numbers $n=1, \dots 4$ and $\mu_s=0$, $\omega^2=0$.
Numerical values of the parameters $\varepsilon^2$, $\mu_d^2$,
the integrals $I_B$ and $\tilde{a}_0$ and the total integrated mass of the solitons are presented in the graphics legend.}
\label{chiralAnglesWN}
\end{center}
\end{figure}

\begin{figure}
\begin{center}
\includegraphics*[scale=0.45,keepaspectratio]{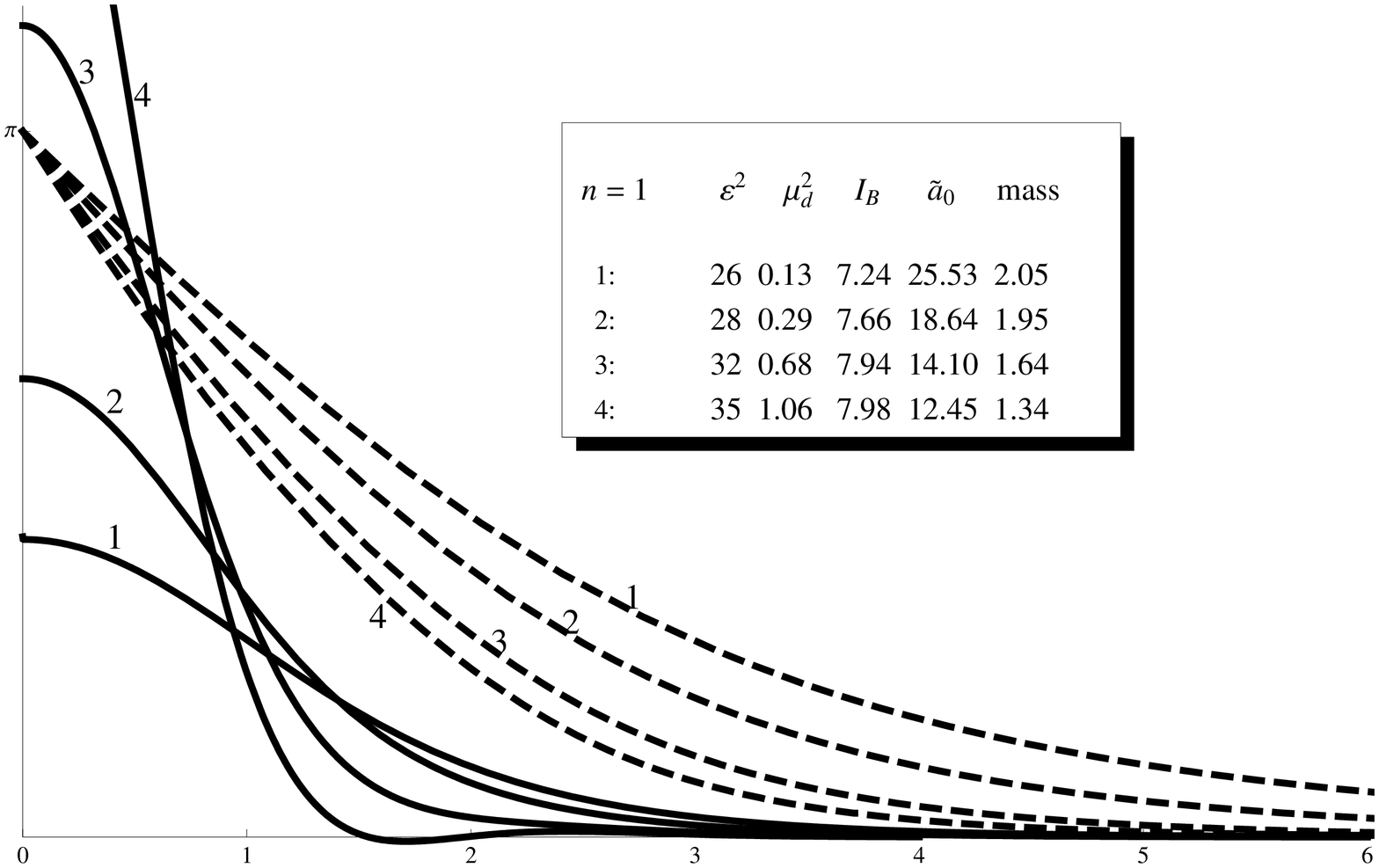}
\caption{The profile functions (dashed lines ) and  the
distribution of the  energy density (solid lines ) of the
$n=1$ solutions with $\mu_s=0$, $\omega^2=0$.
Numerical values of the parameters $\varepsilon^2$, $\mu_d^2$,
the integrals $I_B$ and
$\tilde{a}_0$ and the total integrated mass of the solitons are presented in the graphics legend.}
\label{chiralAnglesWN1}
\end{center}
\end{figure}

\begin{figure}
\begin{center}
\includegraphics*[scale=0.45,keepaspectratio]{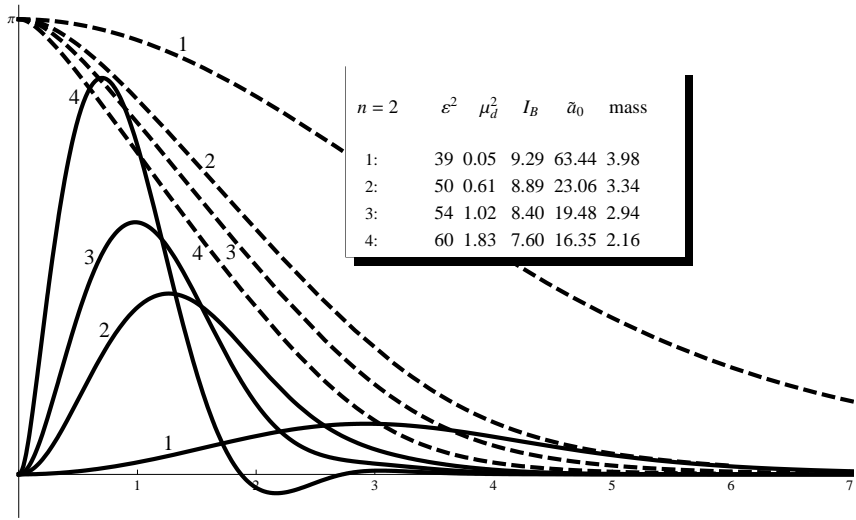}
\caption{The profile functions (dashed lines ) and  the
distribution of the energy density (solid lines ) of the
$n=2$ solutions with $\mu_s=0$, $\omega^2=0$.
Numerical values of the parameters $\varepsilon^2$, $\mu_d^2$,
the integrals $I_B$ and
$\tilde{a}_0$ and the total integrated mass of the solitons are presented in the graphics legend.}
\label{chiralAnglesWN2}
\end{center}
\end{figure}

In figure \ref{chiralAnglesWN}-\ref{chiralAnglesWN2} we presented the results of numerical calculations
for some particular values of the parameters of the model.
The profile functions and the energy density distributions of the soliton solutions with $n=1 \dots 4$
are shown in Fig.~\ref{chiralAnglesWN}. Here we take the value  $\varepsilon^2=30$
in the sector with topological charge $n=1$ and  $\varepsilon^2=40$ for the $n=2$ soliton and
$\varepsilon^2=80$ for the solution solutions with $n=3,4$.

Note that the maxima of the energy density distribution of the $n=2 \dots 4$ solitons
are shifted away from the origin. The size of the solutions increases as the winding number
$n$ increases.

In Figures~\ref{chiralAnglesWN1}, \ref{chiralAnglesWN2} we present the properties of the $\mu_s=0$
soliton solutions within the sectors $n=1$ and $n=2$, respectively. Evidently, the soliton's energy density
distribution becomes more and more spread out, i.e., the characteristic size of the
soliton increases as the parameter $\varepsilon^2$ decreases.

The lines marked with $1$ in both figures represent the solution which corresponds to the
maximal possible value of the dimensionless parameter
$\varepsilon^2$. Our analysis shows that a stable solution does not
exists when this parameter increases beyond this critical value, i.e., quantum correction to the soliton
mass becomes to small to stabilize the solution in the limit of the vanishing potential term. Evidently,
this critical value increases if $\mu_s$ is taking to be non-zero.

On the other hand, increasing of the parameter $\varepsilon^2$ results in increasing
of the values of the momenta of inertia of the quantum soliton. In other words, the quantum correction to the
soliton mass become large and the characteristic size of the configuration decreases. However,
as we can see from plots presented in  Figs.~\ref{chiralAnglesWN1}, \ref{chiralAnglesWN2},
further increasing of the values of the parameter $\varepsilon^2$ yields
negative values of the energy density distribution  (see the curves marked with~$4$), so the corresponding
solutions should be considered rather an artifact of the numerical calculations.

Therefore if $\mu_s=0$, the physical soliton exists for some domain of values of the parameter
$\varepsilon^2$ which is restricted from both sides. Evidently, the quantum correction to
the mass of the soliton cannot be very large, it has to be of order of about a few percent of the
classical mass. The upper bound on the range of values of the parameter   $\varepsilon^2$
corresponds to the ratio $\Delta M /M_{class} \sim 0.4$ which seems to be too large value for a quantum
correction.

\section*{Conclusion}

We investigate the effect of quantization of the rotating baby Skyrmions beyond the usual Bohr-Sommerfeld
framework and the rigid body approximation.
Our results indicate that the canonical quantization of the rotational collective coordinate of the
model may stabilize the soliton solution even in the limit of vanishing potential term.
We have checked that the corresponding radially-symmetric solutions exist in the sectors with
winding numbers $n=1, \dots 4$ for a range of values of the parameters of the model. The energy density
distribution of the quantum baby Skyrmion is exponentially localised as we have seen from
the asymptotic formula \eqref{AsymptoticSolution}. The shape of the quantum soliton depends on the value
of the quantum correction to its mass. At the critical values of the parameters of the
model the configuration becomes unstable or the quantum correction to the mass is too large and the
corresponding energy distribution becomes negative.

It would be interesting to investigate the structure of the quantum
solitons in different topological sectors with various potentials included. In particular,
a question arises about the shape of the minimal energy solution. On the one hand, it is known the choice
of the potential strongly affects the structure of the multisoliton solutions of the model
\cite{Weidig:1998ii,Hen:2007in,Karliner:2009at,Foster:2009vk}. On the other hand,
rotation of the field configuration
may result in the restoration of the radial symmetry, so this problem is worth investigating
further.

This work is supported by the Science Foundation of Ireland
in the framework of the Science Foundation
Ireland (SFI) Research Frontiers Programme (RFP) project RFP07/FPHY330.
We are grateful Tigran Tchrakian for valuable discussions and remarks.

\begin{small}

\end{small}

\end{document}